\documentclass[twocolumn,showpacs,preprintnumbers,amsmath,amssymb]{revtex4}
\usepackage{graphicx}% Include figure files
\usepackage{dcolumn}% Align table columns on decimal point
\usepackage{bm}% bold math
\usepackage{hyperref}

\newcommand{\lap} {\mathcal{L}}
\newcommand{\fur} {\mathcal{F}}
\newcommand{\tlap} {\mathsf{L}}
\newcommand{\tfur} {\mathsf{F}}
\newcommand{\ddp}[2] {\frac{\partial #1}{\partial #2}}

\newcommand{\ddtl}[2] {\frac{\partial^2}{\partial #2^2}#1}
\newcommand{\vt}[1] {\vec{\mathsf{#1}}}
\newcommand{\e}[1] {\mathrm{e}^{#1}}
\newcommand{\st}[1] {\mathsf{#1}}
\newcommand{\rd} {r_\mathrm{ D }}
\newcommand{\vrms} {v_{\mathrm{rms}}}

\newcommand{\pf} {\omega_{\mathrm{p}}}
\newcommand{\pt} {t_{\mathrm{p}}}
\newcommand{\tp} {t_{\mathrm{p}}}
\newcommand{\fd} {f_d}
\newcommand{\re}[1] {\mathrm{#1}}

\begin{document}

\preprint{}

\title{SEMI-ANALYTICAL DESCRIPTION OF THE MODULATOR \\
SECTION OF THE COHERENT ELECTRON 
COOLING}% Force line breaks with \\

\author{Andrey Elizarov $^{1,\,2}$}
\email{andrey.elizarov@stonybrook.edu}
\author{Vladimir Litvinenko $^{1,\,2}$}%
\email{vl@bnl.gov}
 
\affiliation{%
$^1$ Collider-Accelerator Department, Brookhaven National Laboratory, Upton, New York 11973, USA \\
$^2$Department of Physics and Astronomy, Stony Brook University, Stony Brook, New York 11794, USA
}%

\date{\today}% It is always \today, today,

\begin{abstract}
In the coherent electron cooling, the modern hadron beam cooling technique, each hadron receives an individual kick from the electric field of the amplified electron density perturbation created in the modulator by this hadron in a co-propagating electron beam. We developed a method for computing the dynamics of these density perturbations in an infinite electron plasma with any equilibrium velocity distribution -- a possible model for the modulator. We derived analytical expressions for the dynamics of the density perturbations in the Fourier-Laplace domain for a variety of 1D, 2D, and 3D equilibrium distributions of the electron beam. To obtain the space-time dynamics, we employed the fast Fourier transform (FFT) algorithm. We also found an analytical solution in the space-time domain for the 1D Cauchy equilibrium 
distribution, which serves as a benchmark for our general approach based on numerical evaluation of the integral transforms and as a fast alternative to the numerical computations. 
We tested the method for various distributions and initial conditions.

%Valid
%PACS numbers may be entered using the \verb+\pacs{#1}+ command.
%\verb\pacs{52.40.Mj}
%52.27.Aj
\end{abstract}

\maketitle

%\section{\label{sec:level1}First-level heading:\protect\\ The line
%break was forced \lowercase{via} \textbackslash\textbackslash}
\section{\label{intro}Introduction}
A few of years ago, a novel hadron beam cooling technique capable to deal with the accelerators operating in the range of few TeVs, the Coherent electron Cooling (CeC), was proposed \cite{Litvinenko2009}. Currently, a test facility  is under construction at Brookhaven National Laboratory. For the present status of the developments of the machine, we refer to \cite{Pinyaev2012}. The CeC is the modern realization of the stochastic electron cooling, wherein the electron beam serves as a pick-up and a kicker. It records the information about the hadron beam via the electron density perturbations resulting from the shielding of the hadrons.
Then, these perturbations are amplified in the free electron laser (FEL) section, and then, in the kicker, every hadron experience the electric field produced by its own amplified perturbation receiving kicks. Before the kicker, in the dispersion section, each hadron is emplaced such that these kicks accelerate or decelerate it depending on its energy deviation, thereby reducing the energy spread of the hadron beam. 
To analyze the performance of the CeC, all the sections of the device must be studied in detail; in particular, the shielding of a hadron in an electron beam should be computed with high precision.\\

In this article, we offer a theoretical description of the modulator section of the coherent electron cooling, i.e., we address the problem of dynamical shielding of a charged particle in an electron beam. The simplest interpretation of this problem is a screening of a stationary particle in an infinite plasma (by \textit{plasma} we mean a collisionless single-species electron plasma), i.e., the well-known Debye screening. The next step is to consider a moving ion in an infinite plasma. This problem was studied recently and 
the density perturbation for the Lorentz distribution was expressed as a one-dimensional integral \cite{Wang2008}.
The most advanced approach to resolving this problem is a general method for a shielding in a finite electron beam that takes into account the focusing field and the space-charge effects; this methodology was proposed last year by the authors of the present paper \cite{Elizarov2012} and is under development now. There are also simulations of this effect using the PIC ("particle-in-cell") method \cite{Bell}. In this article, we present a solution of this problem via the Fourier and Laplace transforms for the 1D, 2D, and 3D infinite plasmas and various equilibrium distributions. We derived the expressions for the solution in the  Fourier-Laplace space, then inverted them numerically. For the 1D plasma with the Cauchy equilibrium distribution, we found a fully analytical solution in the space-time domain, which gives an opportunity to test the semi-analytical ones involving numerical evaluations of the inverse integral transforms and to perform many-particle computations much faster than with non-exact solutions. This method can also work with empirical equilibrium distributions. As a fast and robust solution, it has its own practical value and it will also serve as a testing ground for the PIC simulations and the general solution mentioned 
for a realistic case of a finite beam.\\

%------------------------------------------------
\section{The Vlasov-Maxwell system} 
%------------------------------------------------
Generally, the shielding of a charged particle in a plasma is described by the Vlasov-Maxwell system of equations \cite{vlasov}, i.e., the dynamics of the electron density is governed by the Vlasov equation and the electro-magnetic field by the Maxwell equations. We first describe the system in a co-moving frame of reference, then derive a formal solution via the integral transforms, then  introduce convenient dimensionless variables, and finally write a solution for a particle moving along a straight line. 
%------------------------------------------------
\subsection{General formulation for an infinite plasma}
%------------------------------------------------
We consider the Vlasov-Maxwell system for the 1D, 2D and 3D plasmas simultaneously, which means that $\vec{x}$ is a one-, two- or three-dimensional vector depending on the dimensionality of the plasma we are considering and by $x$ we denote its absolute value, even for the 1D case; the same conventions are applied for the dimensionless vectors that we will introduce in subsection \ref{dimless}. For the electron phase-space density $f(\vec{x},\vec{p},t)$, the Hamiltonian $H$, and the electric potential $U(\vec{x},t)$, we have:
\begin{align}\label{vlas}
&\ddp{f}{t}+\vec{v}\cdot\ddp{f}{\vec{x}}+\frac{d\vec{p}}{dt}\ddp{f}{\vec{p}}=0,\:\:\:\:\:\: f\equiv f(\vec{x},\vec{p},t),\\
&\vec{v}=\ddp{H}{\vec{p}},\:\:\:\:\:\:\frac{d\vec{p}}{dt}=-\ddp{H}{\vec{x}},\:\:\:\:\:\:H=\frac{p^2}{2m_0}+eU(\vec{x},t),\\
%&\:\:\:\:\:\: H\equiv H(\vec{x},\vec{p},t),\\
&\ddtl{U(\vec{x},t)}{\vec{x}}=-\frac{e}{\epsilon_0}n(\vec{x},t),\:\:\: n(\vec{x},t)=\int f(\vec{x},\vec{p},t)d\vec{p},
\end{align}
and the charge density is $e n(\vec{x})$. For this system, we consider the test charge problem with an external time-dependent density $d(\vec{x},t)$. We assume that $f=f_0+f_1$, where $f_0=f_0(\vec{v})$ is an equilibrium electron density, and $f_1=f_1(\vec{x},\vec{p},t)$ is an unknown perturbation resulting from the interaction with the test charge. The linearized Maxwell-Vlasov system looks as follows:
\begin{align}
&\ddp{f_1}{t}+\vec{v}\cdot\ddp{f_1}{\vec{x}}-\frac{e}{m_0}\ddp{U}{\vec{x}}\ddp{f_1}{\vec{v}}=0,\label{vlas}\\
&\ddtl{U(\vec{x},t)}{\vec{x}}=-\frac{e}{\epsilon_0}\left(n_1(\vec{x},t)+d(\vec{x},t)\right).\label{pois}
\end{align}
%------------------------------------------------
\subsection{Solving the Vlasov-Maxwell system via the integral transforms}
%------------------------------------------------
The Poisson equation (\ref{pois}) can be solved via the Fourier transform:
\begin{align}\label{furim}
k^2\tilde{U}(\vec{k},t)=\frac{e}{\epsilon_0}\left(\tilde{n}_1(\vec{k},t)+\tilde{d}(\vec{k},t)\right),
\end{align}
where $\tilde{U}(\vec{k},t)$, $\tilde{n}_1(\vec{k},t)$, and $\tilde{d}(\vec{k},t)$ are the Fourier images of the corresponding functions.  
Using this solution, we transform the equation (\ref{vlas}) to~\cite{Turski}:
\begin{align}
\tilde{N}_1(\vec{k},s)=\frac{-e^2}{m_0\epsilon_0}\lap\fur_{\vec{k}t}\left(tf_0\left(\vec{v}\right)\right)\left(\tilde{N}_1(\vec{k},s)+\lap\fur d(\vec{x},t)\right),
\end{align} 
where, $\tilde{N}_1(\vec{k},s)$ and $\lap\fur d(\vec{x},t)$ are, respectively, the Laplace-Fourier images of $n_1(\vec{x},t)$ and $d(\vec{x},t)$:
\begin{align}
\tilde{N}_1(\vec{k},s)\equiv\lap\fur n_1(\vec{x},t)=\int
\limits_0^\infty\int n_1(\vec{x},t)\mathrm{e}^{-i\vec{k}\cdot\vec{x}-ts}d\vec{x}dt,
\end{align}
\begin{align} \label{LF1}
\lap\fur d(\vec{x},t)=\int
\limits_0^\infty\int d(\vec{x},t)\mathrm{e}^{-i\vec{k}\cdot\vec{x}-ts}d\vec{x}dt.
\end{align}
We introduce one more notation:
\begin{align}
\label{LF2}
\lap\fur_{\vec{k}t}\left(tf_0\left(\vec{v}\right)\right)=\int\limits_0^\infty\mathrm{e}^{-ts}t\int f_0\left(\vec{v}\right)\mathrm{e}^{-i\vec{k}\cdot\vec{v}t}d\vec{v}dt.
\end{align}
Denoting the inverse Fourier and Laplace transforms, respectively, by $\mathcal{F}^{-1}$ and $\mathcal{L}^{-1}$, we obtain the following expression:
\begin{align}\label{general1}
n_1\left(\vec{x},t\right)=-\frac{e^2}{m_0\epsilon_0}\mathcal{F}^{-1}\mathcal{L}^{-1}\frac{\lap\fur d(\vec{x},t)}{\left(\lap\fur_{\vec{k}t}\left(tf_0\left(\vec{v}\right)\right)\right)^{-1}+\frac{e^2}{m_0\epsilon_0}},
\end{align}
for the details on definitions of the integral transforms and our notations, see Appendix \ref{apena}.
Generally, the expression (\ref{general1}) can be complex. Looking back to our 
initial equations and assuming complex $f_1$, we note that the equation with $\mathrm{Im}f_1$ corresponds to the equation without an external charge, while the equation with $\mathrm{Re}f_1$ is the one 
with it, consequently, $\mathrm{Im}f_1=0$ and it is confirmed by further computations. Hence, the expression (\ref{general1}) is real, as it should be.

Even though the Poisson equations and their Green's functions differ for the 1D, 2D, and 3D cases, their solutions in the Fourier domain (\ref{furim}) and the expression (\ref{general1}) for $n_1\left(\vec{x},t\right)$ have the 
same form.

In proceeding further, we need to specify the dimension of the problem, the external charge density $d(\vec{x},t)$, and the equilibrium distribution, but first we introduce dimensionless variables.
%------------------------------------------------
\subsection{\label{dimless}Introducing dimensionless variables}
%------------------------------------------------
We define the dimensionless variables, denoting them using the sans-serif font, as follows:
\begin{align}
\vt{x}=\frac{\vec{x}}{\rd},\:\:\vt{v}=\frac{\vec{v}}{\vrms},\:\: \st{t}=\frac{t}{\pt},\:\: \vt{k}=\vec{k}\rd,\:\: \st{s}=\frac{s}{\pf},
\end{align}
where 
\begin{align}\label{rms}
&\vrms=\sqrt{\frac{1}{\rho}\int v^2 f_0(\vec{v}) d\vec{v}},\:\:\:\: \pf\equiv\frac{1}{\tp}=\sqrt{\frac{e^2\rho}{m_0\gamma\epsilon_0}},\\
&\rd=\frac{\vrms}{\pf},
\end{align}
are, respectively, the root-mean-square velocity, the plasma frequency, and the Debye radius.
The equilibrium density is normalized via:
\begin{align}
\int f_0(\vec{v})d\vec{v}=\rho.
\end{align}
%For the dimensionless variables variables we always use sans-serif font.
We introduce the dimensionless equilibrium density $\st{f}_0(\vt{v})$ by the relation:
\begin{align}\label{norm}
f_0(\vec{v})=\rho \fd\st{f}_0(\vt{v}),
\end{align}
wherein all the dimensional constants are gathered into $\fd$ and $d$ stands for the dimensionality of the space, and can be $1$, $2,$ or $3$. We have the following dimensionalities for other quantities:
\begin{align}
[\epsilon_0]=\frac{C^2T^2}{L^dM},\:\:\:\:[n(\vec{x},t)]=[\rho]=L^{-d},\:\:\:\:[\fd]=[\vrms]^{-d}.
\end{align}
We note that $\fd$ and $\vrms$ are not the same for the different equilibrium densities and must be computed via (\ref{rms}) and (\ref{norm}); for the non-integrable densities, the values have to be chosen voluntarily; among the densities we consider, only the Cauchy one is of that type. 
Using the dimensionless units, we rewrite formula (\ref{general1}) as follows:
\begin{align}
\st{n}_1\left(\vt{x},\st{t}\right)=-
\label{gen2}
\tlap^{-1}\tfur^{-1}\left[
\frac{\tlap\tfur \big(\st{d}(\vt{x},\st{t})\big)}{\left(\tlap\tfur_{\vt{k}\st{t}}\left(\st{t}\st{f}_0\left(\vt{v}\right)\right)\right)^{-1}\frac{1}{ \fd \vrms^d }+1}\right],
\end{align}
where $\tlap\tfur \big(\st{d}(\vt{x},\st{t})\big)$ and $\tlap\tfur_{\vt{k}\st{t}}\left(\st{t}\st{f}_0\left(\vt{v}\right)\right)$ are the dimensionless analogs of  (\ref{LF1}) and (\ref{LF2}), respectively,  $\frac{1}{ \fd \vrms^d }$ is a dimensionless factor, and $\tlap^{-1}$, $\tfur^{-1}$ are the inverse Laplace and Fourier transforms for the dimensionless variables.
%------------------------------------------------
\subsection{The external point charge}
%------------------------------------------------
We assume that the charge's trajectory is unaffected by the space charge fields and consider 
the charge moving along a straight line $\vec{y}\left(t\right)=\vec{x}_0+\vec{v}_0t$, we have:
\begin{align}
{d}(\vec{x},t)=-Z\delta\left(\vec{x}-\vec{y}\left(t\right)\right),
%\:\:\:\: \rho_{\re{ext}}(\vec{x},t)=-Ze\delta\left(\vec{x}-\vec{y}\left(t\right)\right).
\end{align}
this assumption is reasonable for a hadron moving in an electron beam, as the hadron's mass is much larger than the electron's. For simplicity, we assume $Z=1$ and the final density for the non-unitary charge can be recovered just by multiplying it by $Z$. Using the dimensionless units introduced, for any number of the spatial dimensions, we have:
\begin{align}
\tlap\tfur \big(\st{d}(\vt{x},\st{t})\big)=-\frac{\mathrm{e}^{-i\vt{k}\cdot\vt{x}_0}}{\st{s}+i\vt{k}\cdot\vt{v}_0},
\:\:\:\: \vt{y}(\st{t})=\vt{x}_0+\vt{v}_0\st{t}.
\end{align}
Finally, we can write the expression for the electron density perturbation resulting from the interaction with the external charge moving along a straight line $\vt{y}\left(\st{t}\right)=\vt{x}_0+\vt{v}_0\st{t}$, valid in 1D, 2D, and 3D spaces:
\begin{align}\label{genfin}
\st{n}_1\left(\vt{x},\st{t}\right)=
\tlap^{-1}\tfur^{-1}\left[
\frac{\mathrm{e}^{-i\vt{k}\cdot\vt{x}_0}}{\left(\frac{\fd^{-1} \vrms^{-d}}{\tlap\tfur_{\vt{k}\st{t}}\left(\st{t}\st{f}_0\left(\vt{v}\right)\right)}+1\right)\left(\st{s}+i\vt{k}\cdot\vt{v}_0\right)}\right].
\end{align}
In the next section, we consider this solution for some particular equilibrium densities $\st{f}_0\left(\vt{v}\right)$; for each case, we just need to compute $\tlap\tfur_{\vt{k}\st{t}}\left(\st{t}\st{f}_0\left(\vt{v}\right)\right)$ and $\fd^{-1} \vrms^{-d}$ and insert them into (\ref{genfin}).
%------------------------------------------------
\section{Application to the particular equilibrium distributions}
%------------------------------------------------
Generally, the equilibrium distribution $\st{f}_0\left(\vt{v}\right)$ has to be a solution of the unperturbed Vlasov equation, i.e., it has to be a 
function of the unperturbed Hamiltonian, in our dimensionless units it is $\st{v}^2$, thus we consider the following functions:
\begin{align}
\delta(\st{v}^2-1),\:\:
\Theta(-\st{v}^2+1),\:\:
\e{-\st{v}^2},\:\:
(1+\st{v}^2)^{-\frac{1+d}{2}}.
\end{align}
They correspond to the Kapchinskij-Vladimirskij (KV), water-bag (WB), normal (or Maxwell), and Cauchy (or Lorentz) equilibrium distributions, $\Theta(v)$ stands for the Heaviside step function, $d$ is the dimensionality of the space, and $\vt{v}$ is a 
one-, two-, or three-dimensional vector. 
However, for the case of an infinite plasma that we are considering here, any function of velocity is a solution of the unperturbed Vlasov equation; thus, all our formulae can be easily generalized for the equilibrium distributions of the form:
\begin{align}
\st{f}_0\left(\sum\limits_{i=1}^d(\st{a}_i\st{v}_i)^2\right),
\end{align}
corresponding to an anisotropic plasma, where $\st{a}_i$, $i=1,\,...,\,d$ are dimensionless constants characterizing the plasma's temperatures. 
The changes should be applied only to the expression for $\tlap\tfur_{\vt{k}\st{t}}\left(\st{t}\st{f}_0\left(\vt{v}\right)\right)$, i.e., $\st{k}_i$ should be substituted with $\st{k}_i/\st{a}_i$ for $i=1,\,...,\,d$, and the whole expression should be divided by $\prod_{i=1}^d\st{a}_i$.
\\

For the 1D Cauchy distribution, the inverse Laplace and Fourier transforms in (\ref{genfin}) can be evaluated analytically, while for the other distributions, the numerical techniques should be applied. We describe the 1D Cauchy case in detail and just quote the results for the other distributions starting with the KV and WB, which have different expressions for $\tlap\tfur_{\vt{k}\st{t}}\left(\st{t}\st{f}_0\left(\vt{v}\right)\right)$ in spaces of different dimensionalities. We conclude this section with the Cauchy and Maxwell distributions that have the same expressions for this quantity in all cases. Although, below we present $\vrms$ computed via (\ref{rms}), we considered dimensionless equilibrium distributions, $\st{f}_0\left(\vt{v}\right)$, corresponding to
\begin{align}
\vrms=\sqrt{\frac{H_c}{\beta}},%\text{, and }\frac{1}{ \fd \vrms }=1,
\end{align}
with this $\vrms$, in all cases, $\frac{1}{ \fd \vrms^d }=1$ and $\st{f}_0\left(\vt{v}\right)$ has a simpler form.
%------------------------------------------------
\subsection{The 1D Cauchy distribution}
%------------------------------------------------
 Here, we assume  $f_0\left(\vec{v}\right)=\rho\frac{\beta}{H_c}\frac{1}{\pi\left(1+\frac{\beta}{H_c}v^2\right)}$, where $\beta$ and $H_c$ are dimensional constants that can be used for fitting the experimental distributions. Computing $\lap\fur_{\vec{k}t}\left(tf_0\left(\vec{v}\right)\right)$ via (\ref{LF2}), we obtain the following expression:
\begin{align}
\lap\fur_{\vec{k}t}\left(tf_0\left(\vec{v}\right)\right)=\frac{\rho}{\left(s+k\sqrt{\frac{H_c}{\beta}}\right)^2},
\end{align}
or, using the dimensionless variables:
\begin{align}
&\st{f}_0(\vt{v})=\frac{1}{\pi\left(1+\st{v}^2\right)},\:\:\:\:\:\:
\tlap\tfur_{\vt{k}\st{t}}\left(\st{t} \st{f}_0\left(\vt{v}\right)\right)=\frac{1}{(\st{s}+\st{k})^2},\\
&\vrms=\sqrt{\frac{H_c}{\beta}},\:\:\:\: \fd=\sqrt{\frac{\beta}{H_c}},\:\:\:\:  \frac{1}{ \fd \vrms^d }=1.
\end{align}
Then, we insert the expression for $\tlap\tfur_{\vt{k}\st{t}}\left(\st{t} \st{f}_0\left(\vt{v}\right)\right)$ into the formula (\ref{genfin}) and obtain:
\begin{align}
\st{n}_1\left(\vt{x},\st{t}\right)=\tlap^{-1}\tfur^{-1}\left[
\frac{\mathrm{e}^{-i\vt{k}\cdot\vt{x}_0}}{\left(1+(\st{s}+\st{k})^2\right)\big(\st{s}+i\vt{k}\cdot\vt{v}_0\big)}\right].
\end{align}
For all distributions we are considering, excepting the 1D Cauchy, the inverse integral transforms in the corresponding expressions for $\st{n}_1\left(\vt{x},\st{t}\right)$  have to be inverted numerically, while, for the 1D Cauchy, they 
can be computed analytically giving the following expression:
\begin{align}\label{huge3}
\st{n}_1\left(\vt{x},\st{t}\right)=\frac{1}{4\pi}\frac{1}{\st{v}_0-i}\left(\e{-\mathcal{A}_+}\left(\re{Ei}(\mathcal{A}_+)-\re{Ei}(\mathcal{B}_+)\right)+\notag\right. \\ \left.+
\e{\mathcal{A}_+}\left(\re{E}_1(\mathcal{A}_+)-\re{E}_1(\mathcal{B}_+)\right)\right)+\notag \\+
\frac{1}{4\pi}\frac{1}{\st{v}_0+i}\left(\e{-\mathcal{A}_-}\left(\re{Ei}(\mathcal{A}_-)-\re{Ei}(\mathcal{B}_-)\right)+\right.\notag \\ \left.+
\e{\mathcal{A}_-}\left(\re{E}_1(\mathcal{A}_-)-\re{E}_1(\mathcal{B}_-)\right)\right),
\end{align}
where
\begin{align}
\mathcal{A}_\pm=\frac{\st{t}\st{v}_0-\st{x}+\st{x}_0}{1\pm i\st{v}_0},\:\:\:\: \mathcal{B}_\pm=\frac{\st{x}_0-\st{x}\pm i\st{t}}{1\pm i\st{v}_0},
\end{align}
and $\re{E}_1(z)$ and $\re{Ei}(z)$ are the exponential integral functions \cite{as} that can be computed via the series expansions, for details, see Appendix \ref{eiig}. The whole expression (\ref{huge3}) is real even though it contains complex numbers.\\ 

\begin{figure}[h]
\includegraphics[scale=0.42]{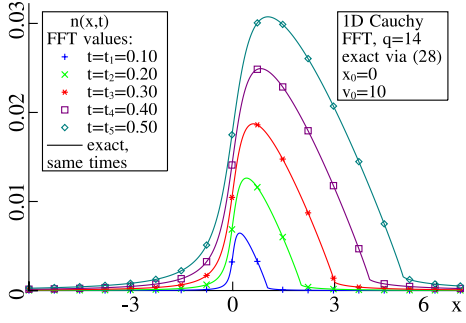}
\includegraphics[scale=0.42]{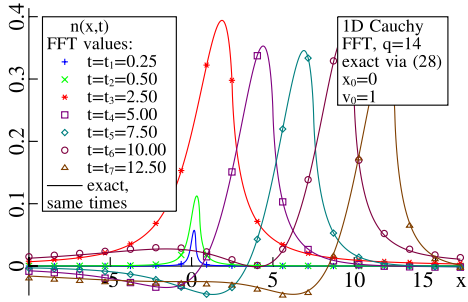}
\includegraphics[scale=0.42]{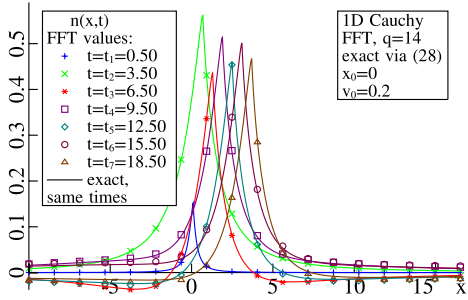}
\caption{\label{comp}The exact values and the ones obtained numerically via the FFT for the 1D Cauchy distribution, for the various velocities of the external charge. Solid lines represent the exact values and the FFT values are shown by marks of different shapes regarding the times.}
\end{figure}

In Fig. \ref{comp}, we show
the densities obtained via the exact formula (\ref{huge3}) and the ones obtained by the discussed in section \ref{num} numerical inversion of the integral transforms for $\st{x}_0=0$ and $\st{v}_0=0.2,\,1.0,\,10.0$. 
We note the perfect agreement of the exact solution and the numerical one. The solution has several interesting features, i.e., starting from some time, the left tail of the density has negative values, meaning that there is an accumulation of the charge of the same sign as that of the external perturbation, its maximum is oscillating, and the shape of the peak depends on the charge's velocity, being spiky for small velocities, widening as it increases, and, for large velocities, a discontinuity of the density's shape derivative appears in the right tail. All these features are equally well captured by the numerical computations and the analytical formula. We will comment further on the parameters' values in subsection \ref{res}.
%ed in previous 
%works, as time ranges considered were much smaller.
 %------------------------------------------------
\subsection{The 1D KV and WB distributions}
%------------------------------------------------
For the 1D KV distribution, we have  $f_0(\vec{v})=\rho H_c \sqrt{\frac{\beta}{H_c}}\delta\left(\beta v^2-H_c\right)$. Using the dimensionless variables, we obtain:
\begin{align}
&\st{f}_0(\vt{v})=\delta(\st{v}^2-1),\:\:\:
\tlap\tfur_{\vt{k}\st{t}}\left(\st{t} \st{f}_0\left(\vt{v}\right)\right)=\frac{\st{s}^2-\st{k}^2}{\left(\st{s}^2+\st{k}^2\right)^2},
\end{align}
\begin{align}
&\vrms=\sqrt{\frac{H_c}{\beta}},\:\:\:\: \fd=\sqrt{\frac{\beta}{H_c}},\:\:\:\:  \frac{1}{ \fd \vrms}=1.
%&\st{n}_1\left(\vt{x},\st{t}\right)=\tlap^{-1}\tfur^{-1}\left[
%\frac{\mathrm{e}^{-i\vt{k}\cdot\vt{x}_0}}{\left(\frac{\left(\st{s}^2+\st{k}^2\right)^2}{\st{s}^2-\st{k}^2}+1\right)\left(\st{s}+i\vt{k}\cdot\vt{v}_0\right)}\right]
\end{align}
For  the 1D WB, we have:
\begin{align}
&f_0\left(v\right)=\frac{1}{2}\rho\sqrt{\frac{\beta}{ H_c}} \Theta\left(-\frac{\beta}{H_c} v^2+1\right),
\end{align}
\begin{align}
&\st{f}_0(\vt{v})=\frac{1}{2}\Theta(1-\st{v}^2),\:\:\:\: \tlap\tfur_{\vt{k}\st{t}}\left(\st{t} \st{f}_0\left(\vt{v}\right)\right)=\frac{1}{\st{k}^2+\st{s}^2},
\end{align}
\begin{align}
&\vrms=\frac{1}{\sqrt{3}}\sqrt{\frac{H_c}{\beta}},\:\:\:\: \fd=\sqrt{\frac{\beta}{H_c}},\:\:\:\:  \frac{1}{ \fd \vrms }=\sqrt{3}.
%&\st{n}_1\left(\vt{x},\st{t}\right)=\tlap^{-1}\tfur^{-1}\left[
%\frac{\mathrm{e}^{-i\vt{k}\cdot\vt{x}_0}}{\left(1+\sqrt{3}(\st{k}^2+\st{s}^2)\right)\big(\st{s}+i\vt{k}\cdot\vt{v}_0\big)}\right]
\end{align}
 %------------------------------------------------
\subsection{The 2D KV and WB distributions}
%------------------------------------------------
In the 2D case, we have for the KV distribution: 
\begin{align}
&f_0\left(\vec{v}\right)=\rho H_c\frac{1}{\pi}\frac{\beta}{H_c}\delta\left(\beta v^2-H_c\right),
\end{align}
\begin{align}
&\st{f}_0(\vt{v})=\frac{1}{\pi}\delta\left(\st{v}^2-1\right),\:\:\:
\tlap\tfur_{\vt{k}\st{t}}\left(\st{t} \st{f}_0\left(\vt{v}\right)\right)=\frac{\st{s}}{\left(\st{s}^2+\st{k}^2\right)^\frac{3}{2}},
\end{align}
\begin{align}
&\vrms=\sqrt{\frac{H_c}{\beta}},\:\:\:\: \fd=\frac{\beta}{H_c},\:\:\:\:  \frac{1}{ \fd \vrms^2 }=1,
%&\st{n}_1\left(\vt{x},\st{t}\right)=\tlap^{-1}\tfur^{-1}\left[
%\frac{\mathrm{e}^{-i\vt{k}\cdot\vt{x}_0}}{\left(1+\frac{1}{\st{s}}\left(\st{s}^2+\st{k}^2\right)^\frac{3}{2}\right)\big(\st{s}+i\vt{k}\cdot\vt{v}_0\big)}\right].
\end{align}
and, for the 2D WB:
\begin{align}
&f_0\left(\vec{v}\right)=\rho\frac{\beta}{Hc}\frac{1}{\pi}\Theta\left(-\frac{\beta}{H_c} v^2+1\right),
\end{align}
\begin{align}
&\st{f}_0(\vt{v})=\frac{1}{\pi}\Theta\left(1-\st{v}^2\right),\:\:\:
\tlap\tfur_{\vt{k}\st{t}}\left(\st{t} \st{f}_0\left(\vt{v}\right)\right)=\frac{2}{\st{k}^2}\frac{\sqrt{\st{k}^2+\st{s}^2}-s}{\sqrt{\st{k}^2+\st{s}^2}},
\end{align}
\begin{align}
&\vrms=\frac{1}{\sqrt{2}}\sqrt{\frac{H_c}{\beta}},\:\:\:\: \fd=\frac{\beta}{H_c},\:\:\:\:  \frac{1}{ \fd \vrms^2 }=2.
\end{align}
 %------------------------------------------------
\subsection{The 3D KV and WB distributions}
%------------------------------------------------
The expressions are slightly bulkier in the 3D case; we obtain for the 3D KV:
\begin{align}
&f_0\left(\vec{v}\right)=\rho\delta\left(\beta v^2-H_c\right)\frac{1}{2\pi}H_c\left(\frac{\beta}{H_c}\right)^{3/2},\\
&\st{f}_0(\vt{v})=\frac{1}{2\pi}\delta(\st{v}^2-1),\:\:\:\tlap\tfur_{\vt{k}\st{t}}\left(\st{t} \st{f}_0\left(\vt{v}\right)\right)=\frac{1}{2}
\hat{I}(\st{s},\vt{k},1),\\
&\vrms=\sqrt{\frac{H_c}{\beta}},\:\:\:\: \fd=\left(\frac{\beta}{H_c}\right)^{\frac{3}{2}},\:\:\:\:  \frac{1}{ \fd \vrms^3 }=1,
\end{align}
where 
\begin{align}
\hat{I}(\st{s},\vt{k},\st{v})=\frac{i\st{k}_3\st{v}\left(\mathcal{S}_--\mathcal{S}_+\right)
+s\left(\mathcal{S}_-+\mathcal{S}_+\right) }{(\st{s}^2+\st{k}^2\st{v}^2)\mathcal{S}_-\mathcal{S}_+},\\
\mathcal{S}_\pm=\sqrt{\left(\st{s}\pm i\frac{\st{k}_3^2\st{v}}{\st{k}}\right)^2},
\end{align}
and $\st{k}_3$ is a third component of $\vt{k}$; for the 3D WB, we obtain:
\begin{align}
&f_0\left(\vec{v}\right)=\rho\frac{3}{4\pi}\Theta\left(-\frac{\beta}{H_c} v^2+1\right)\left(\frac{\beta}{H_c}\right)^{3/2},\\
&\st{f}_0(\vt{v})=\frac{3}{4\pi}\Theta(1-\st{v}^2),\\ \label{3dwb}
&\tlap\tfur_{\vt{k}\st{t}}\left(\st{t} \st{f}_0\left(\vt{v}\right)\right)=\frac{3}{2}\int\limits_0^1\st{v}^2
\hat{I}(\st{s},\vt{k},\st{v})d\st{v},\\
&\vrms=\sqrt{\frac{3H_c}{5\beta}},\:\:\: \fd=\left(\frac{\beta}{H_c}\right)^{\frac{3}{2}},\:\:\:  \frac{1}{ \fd \vrms^3 }=\left(\frac{5}{3}\right)^\frac{3}{2},
\end{align}
the integral in (\ref{3dwb}) has to be computed numerically.

 %------------------------------------------------
\subsection{The general Cauchy distribution}
%------------------------------------------------
For the Cauchy distribution, we obtained the expressions valid in 1D, 2D, and 3D cases:

\begin{align}
&f_0\left(\vec{v}\right)=\rho\left(\frac{\beta}{H_c}\right)^{\frac{d}{2}}\frac{\Gamma(\frac{1+d}{2})}{\Gamma(\frac{1}{2})\pi^{\frac{d}{2}}}\left(1+\frac{\beta v^2}{H_c}\right)^{-\frac{1+d}{2}},\\
&\st{f}_0(\vt{v})=\frac{\Gamma(\frac{1+d}{2})}{\Gamma(\frac{1}{2})\pi^{\frac{d}{2}}}\frac{1}{\left(1+\st{v}^2\right)^{\frac{1+d}{2}}}
,\:
\tlap\tfur_{\vt{k}\st{t}}\left(\st{t} \st{f}_0\left(\vt{v}\right)\right)=\frac{1}{\left(\st{s}+\st{k}\right)^{2}},\\&
\vrms=\sqrt{\frac{H_c}{\beta}},\:\:\:\: \fd=\left(\frac{\beta}{H_c}\right)^{\frac{d}{2}},\:\:\:\:  \fd^{-1} \vrms^{-d}=1.
\end{align}
 %------------------------------------------------
\subsection{The normal distribution}
%------------------------------------------------
For the Maxwell distribution, we also found universal formulas valid in 1D, 2D, and 3D cases:
\begin{figure*}
\includegraphics[scale=0.40]{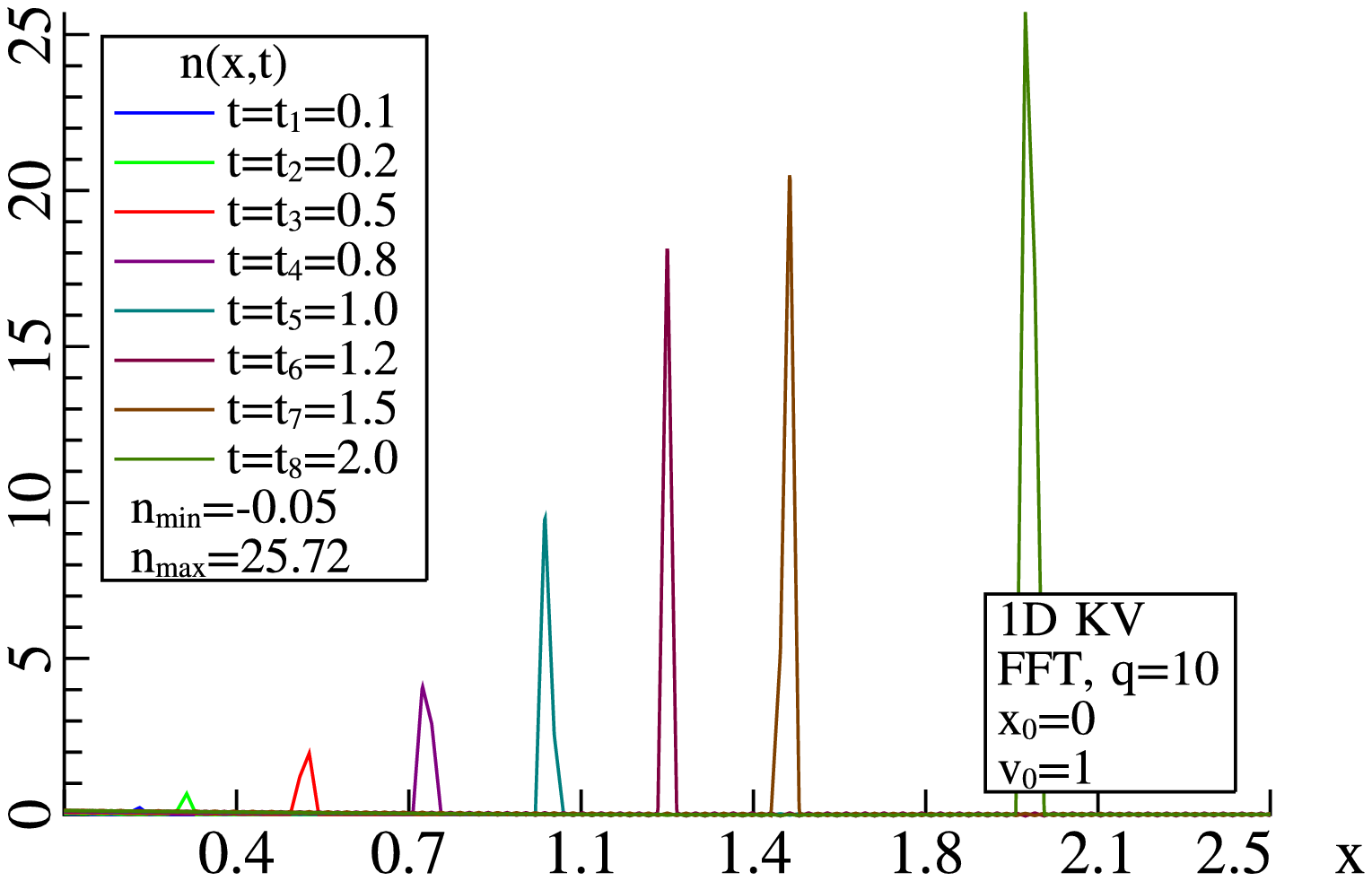}
\includegraphics[scale=0.40]{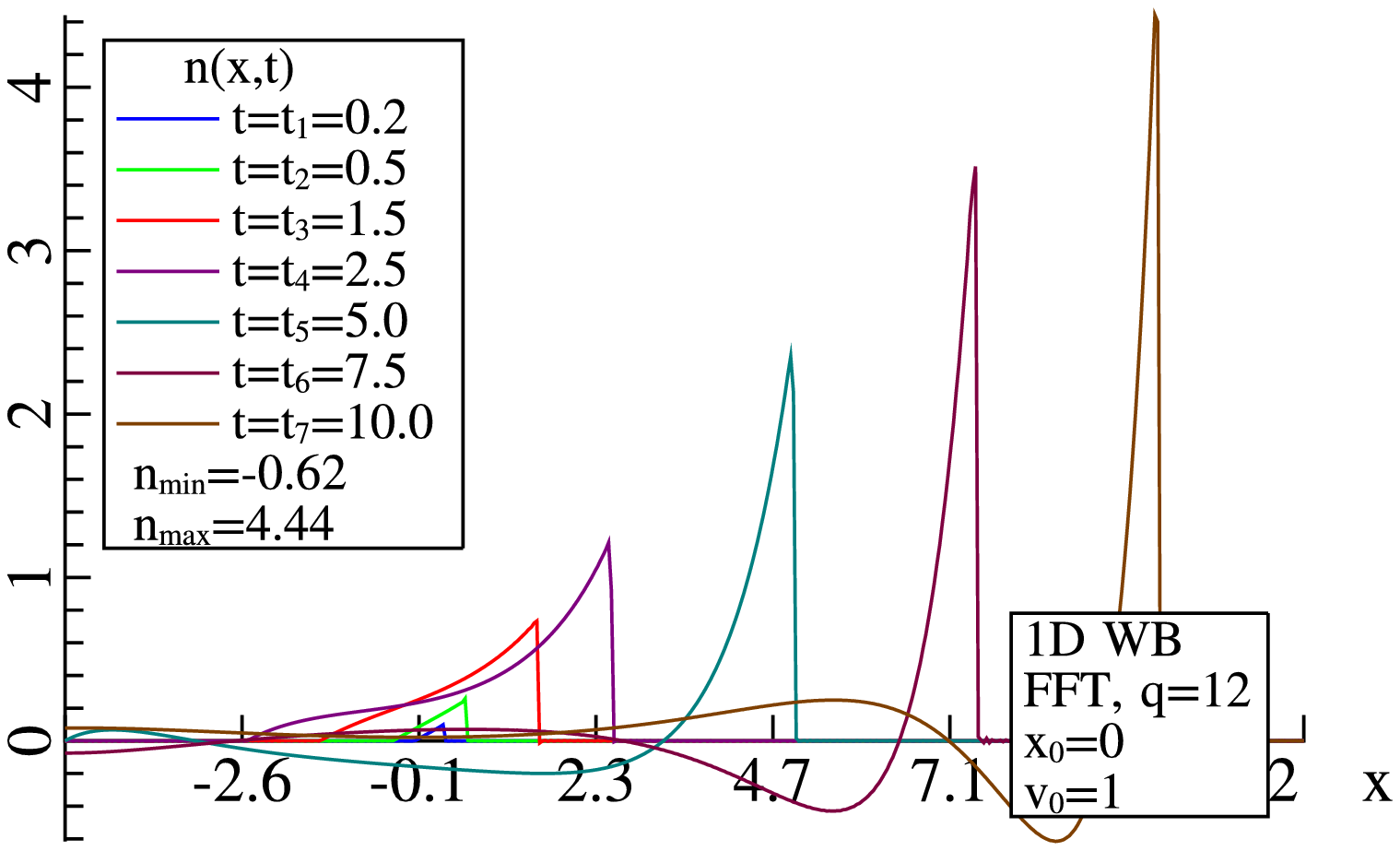}
\includegraphics[scale=0.40]{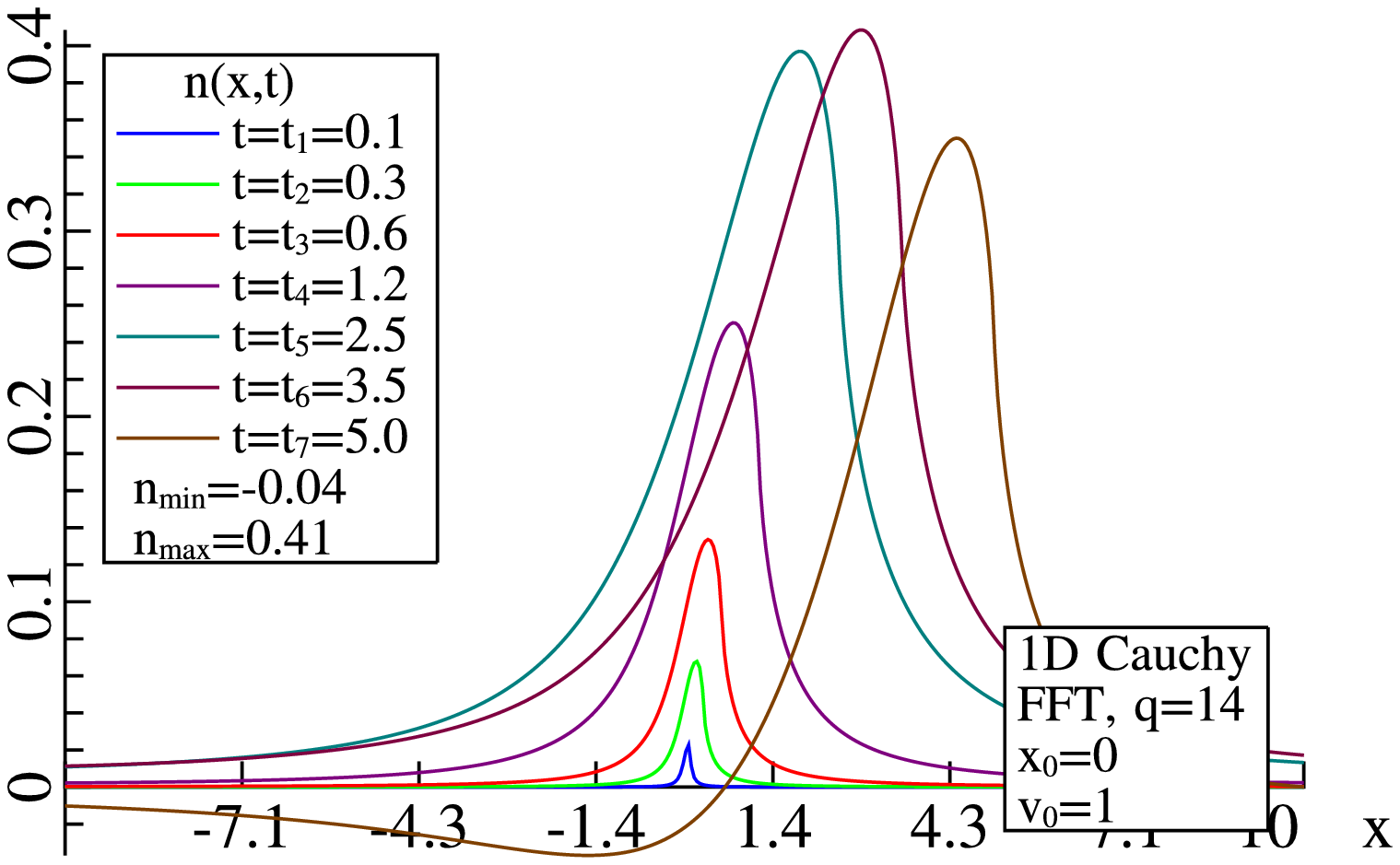}
\includegraphics[scale=0.40]{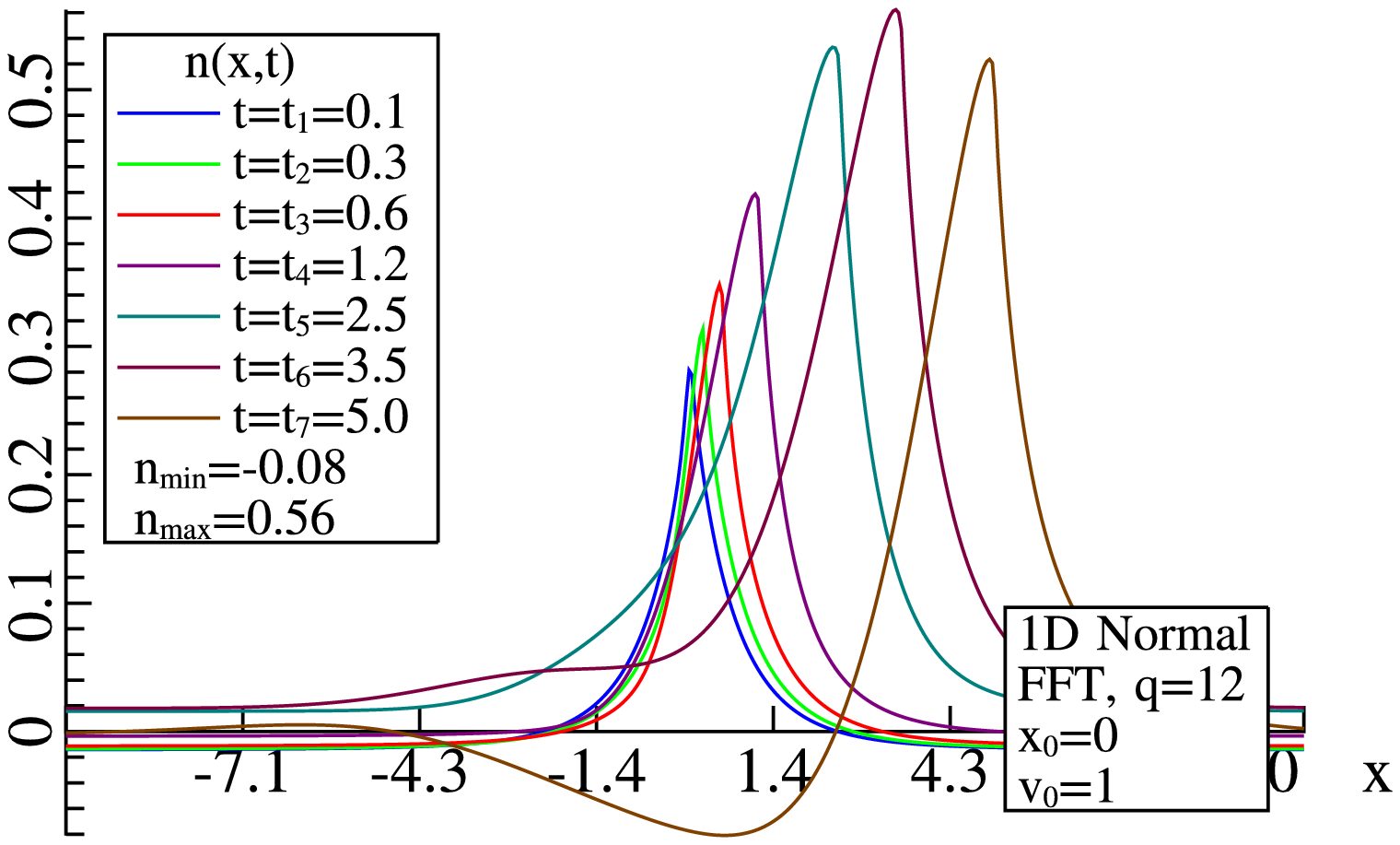}
\caption{\label{1d}The density $n(\vt{x},\st{t})$ for the KV, WB, normal, and Cauchy distributions in 1D space.}
\end{figure*}
\begin{align}
&f_0\left(\vec{v}\right)=\frac{\rho}{\pi^{d/2}}\left(\frac{H_c}{\beta}\right)^{-\frac{d}{2}}\mathrm{exp}^{-\frac{\beta v^2}{H_c}},\:\:\:\: \st{f}_0(\vt{v})=\pi^{-\frac{d}{2}}\e{-\st{v}^2},\\
&\tlap\tfur_{\vt{k}\st{t}}\left(\st{t} \st{f}_0\left(\vt{v}\right)\right)=\frac{2}{\st{k}^2}
\left[1-\sqrt{\pi}\e{\frac{\st{s}^2}{\st{k}^2}}\frac{s}{\st{k}}\re{Erfc}\frac{s}{\st{k}}
\right],\\
&\vrms=\sqrt{\frac{dH_c}{2\beta}},\, \fd=\left(\frac{\beta}{H_c}\right)^{\frac{d}{2}},\,\fd^{-1} \vrms^{-d}=\left(2/d\right)^{\frac{d}{2}},
\end{align}
where $\re{Erfc}(z)$ is the complementary error function \cite{as}, for its definition and some computational details, see Appendix \ref{erf}.
 \begin{figure}
\includegraphics[scale=0.40]{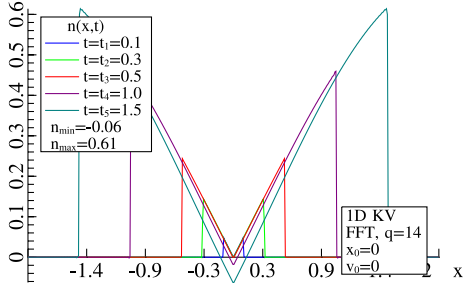}
\caption{\label{1dkv}The density $n(\vt{x},\st{t})$ for the 1D KV for $\st{v}_0=0$.}
\end{figure}
 \begin{figure*}
\includegraphics[scale=0.40]{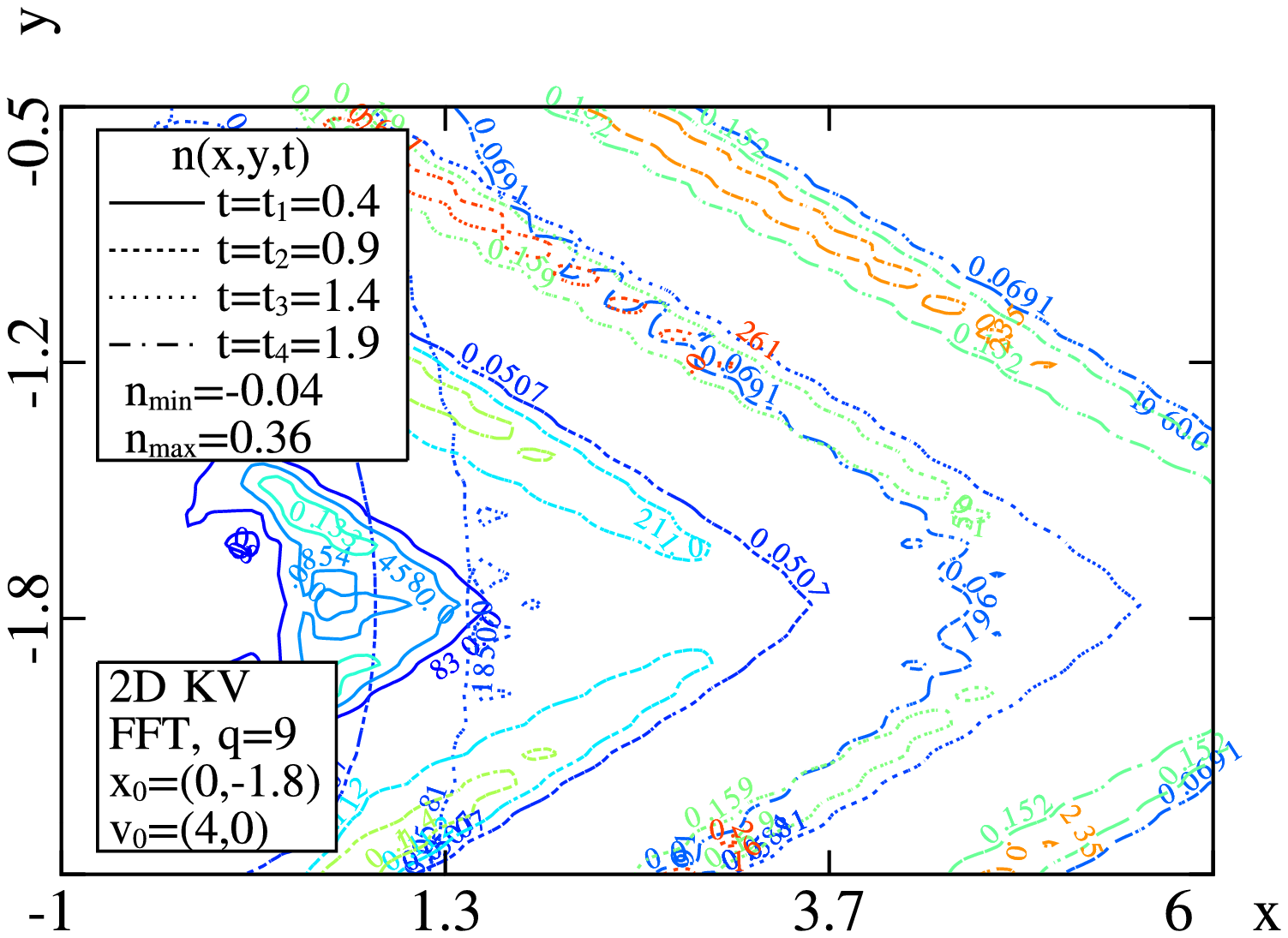}
\includegraphics[scale=0.40]{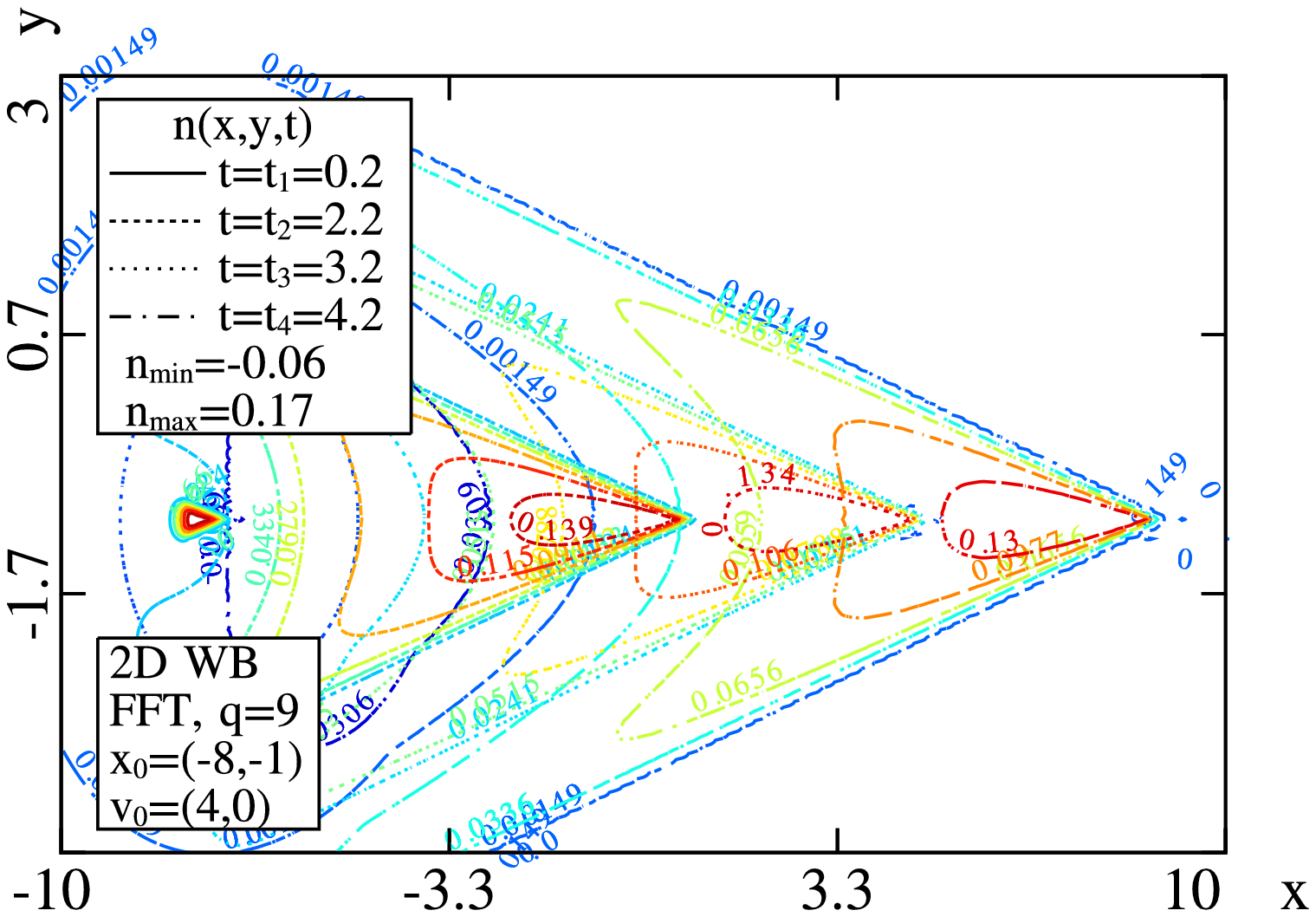}
\includegraphics[scale=0.40]{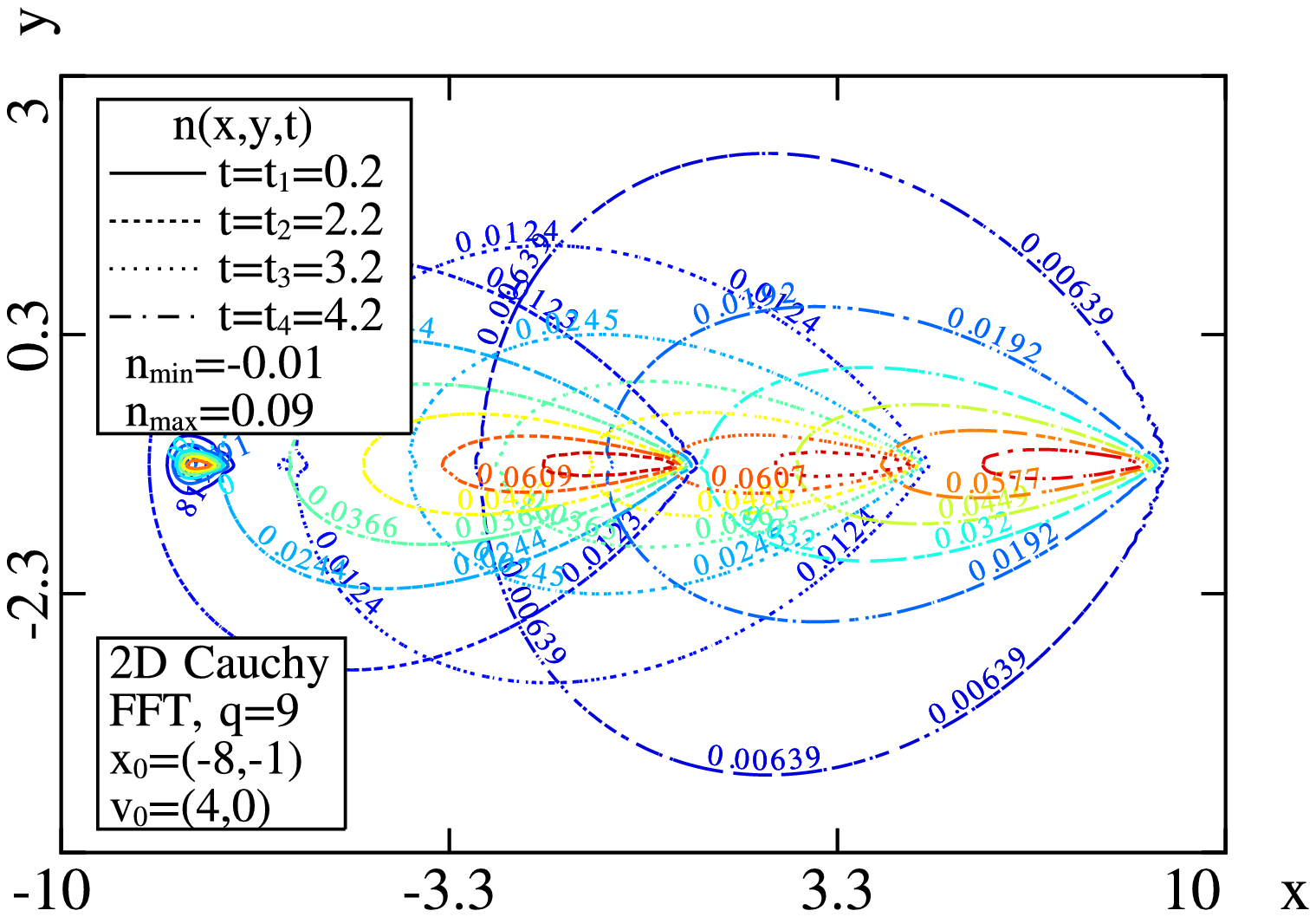}
\includegraphics[scale=0.40]{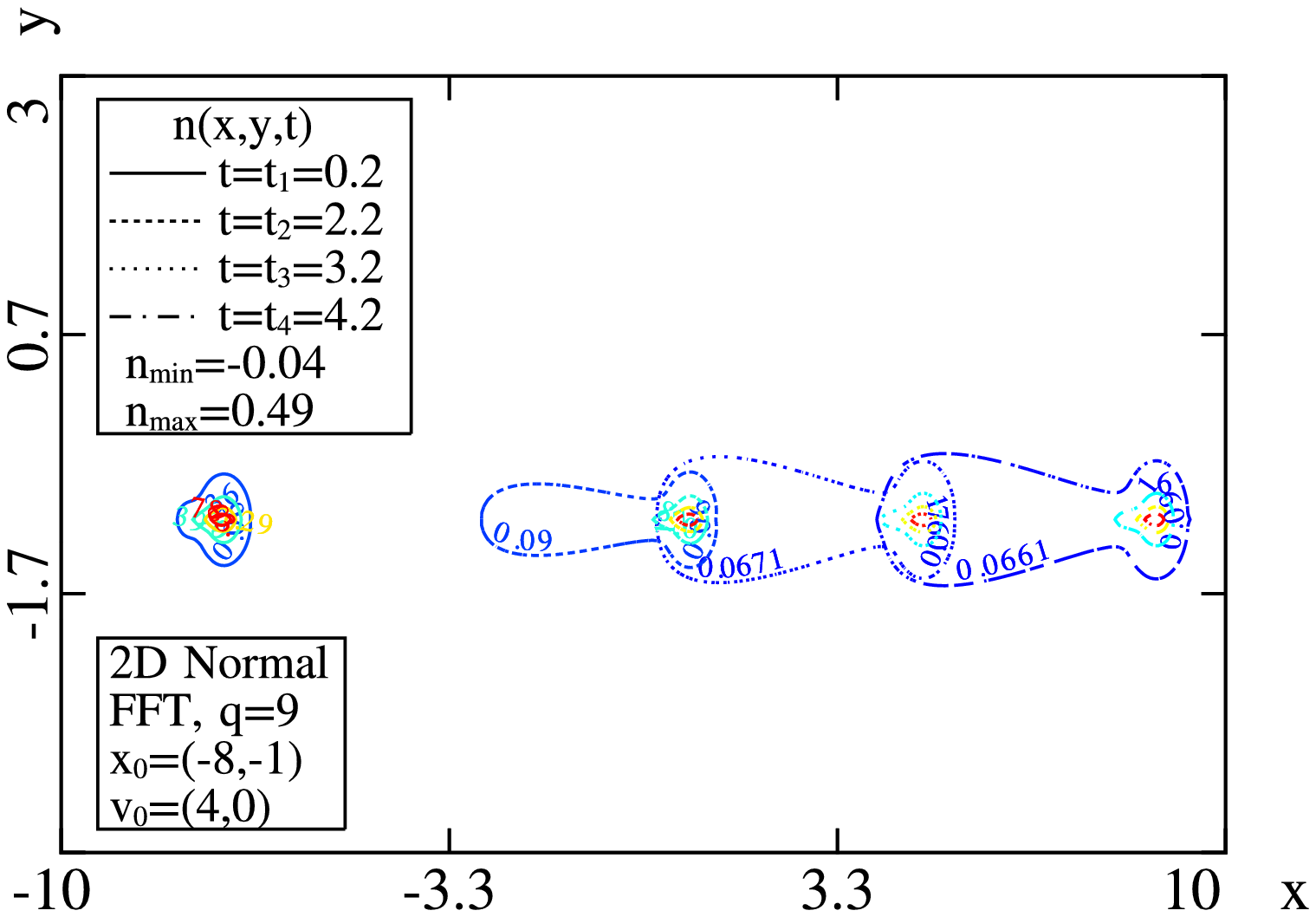}
\caption{\label{2d}The density $n(\vt{x},\st{t})$ for the KV, WB, normal, and Cauchy distributions in 2D space.}
\end{figure*}
 \begin{figure*}
\includegraphics[scale=0.420]{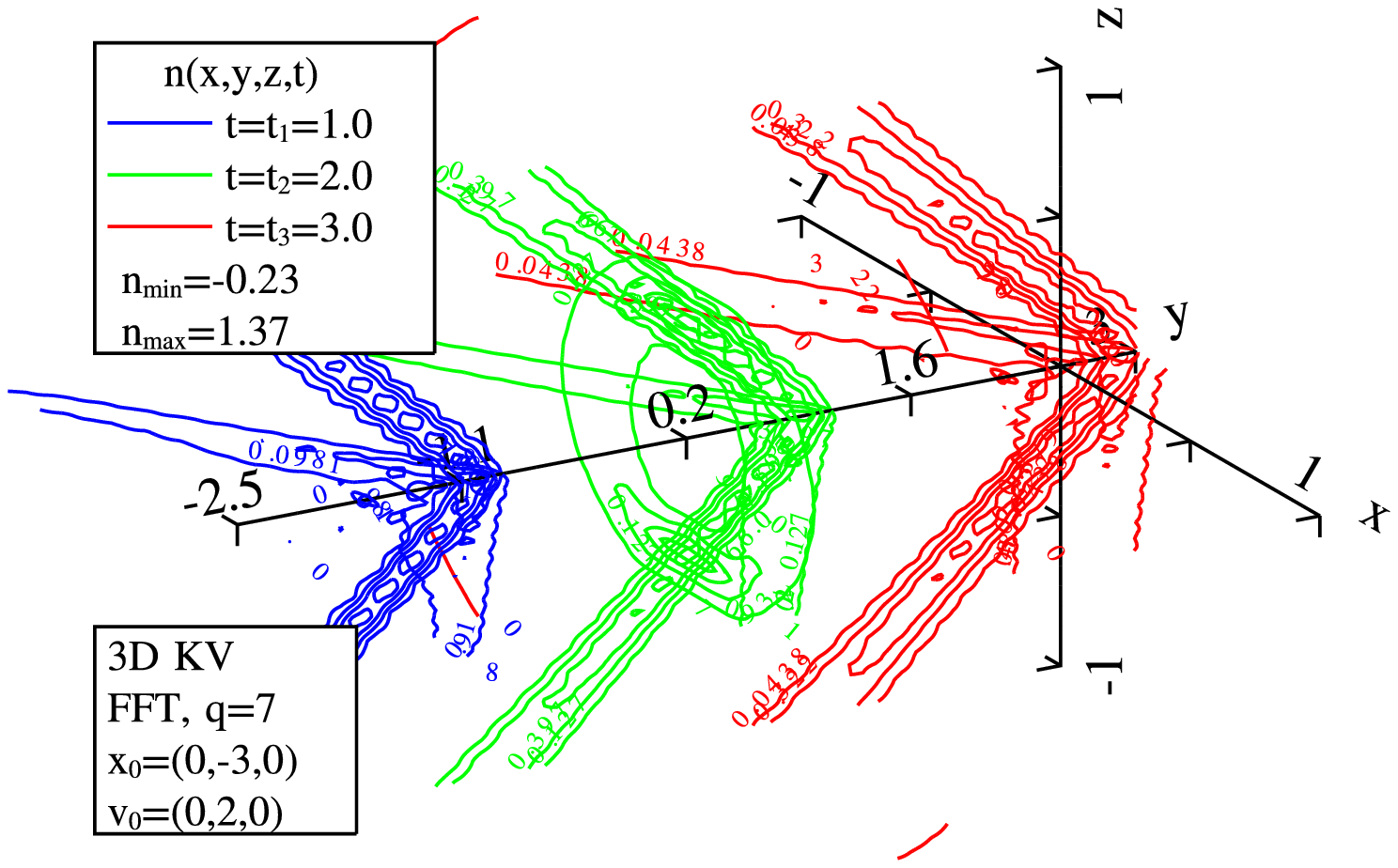}
\includegraphics[scale=0.420]{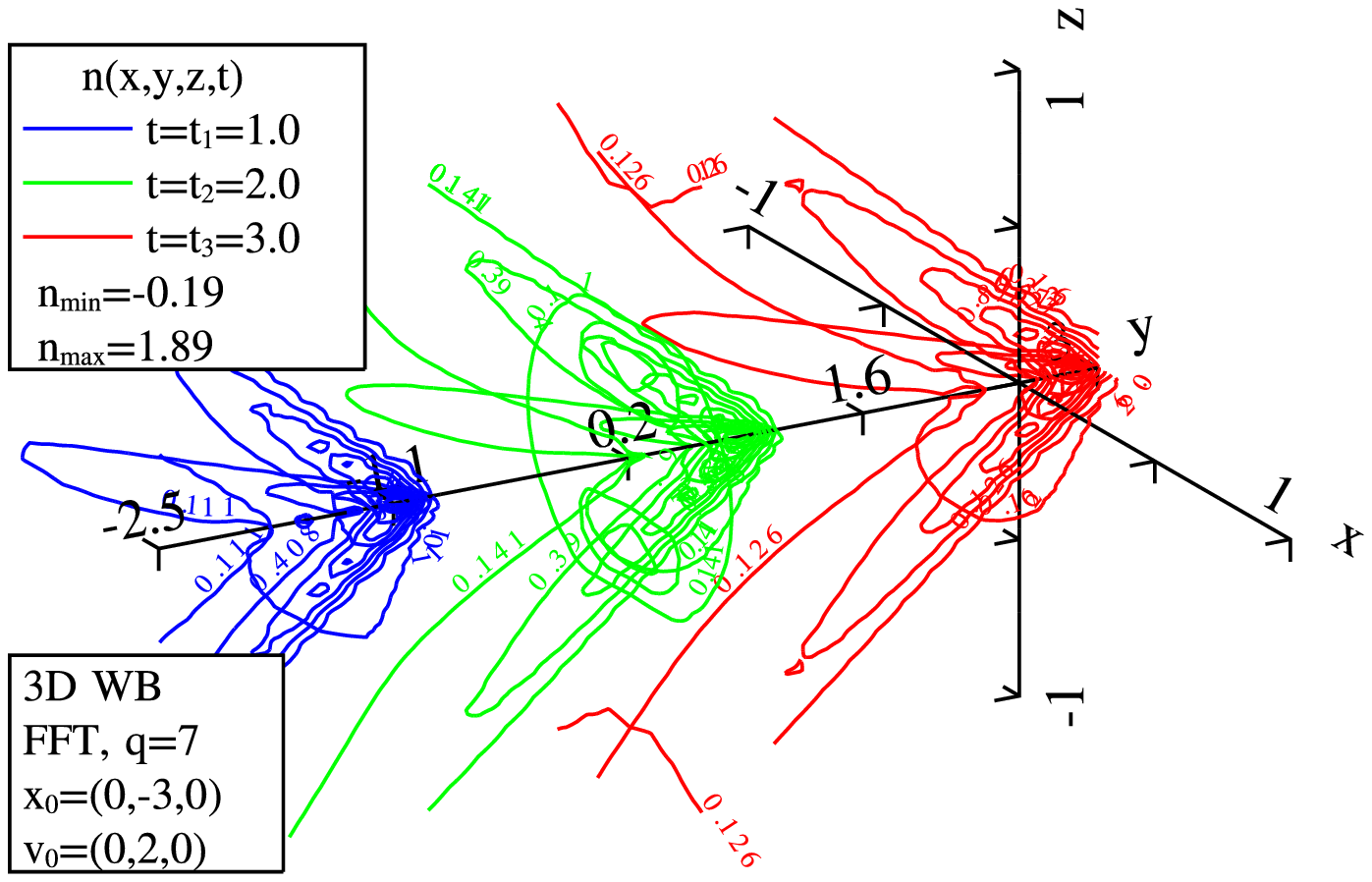}
\includegraphics[scale=0.420]{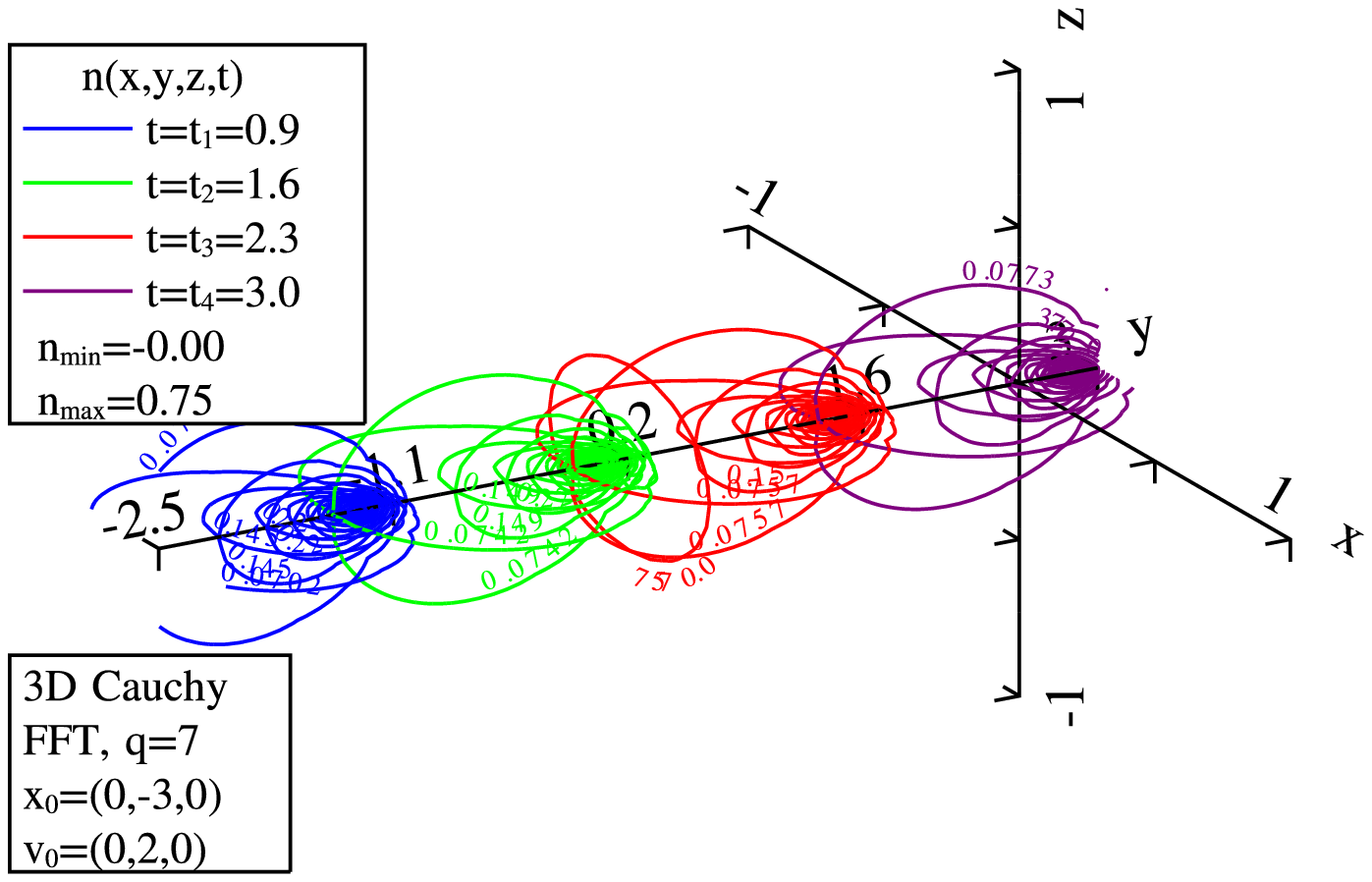}
\includegraphics[scale=0.420]{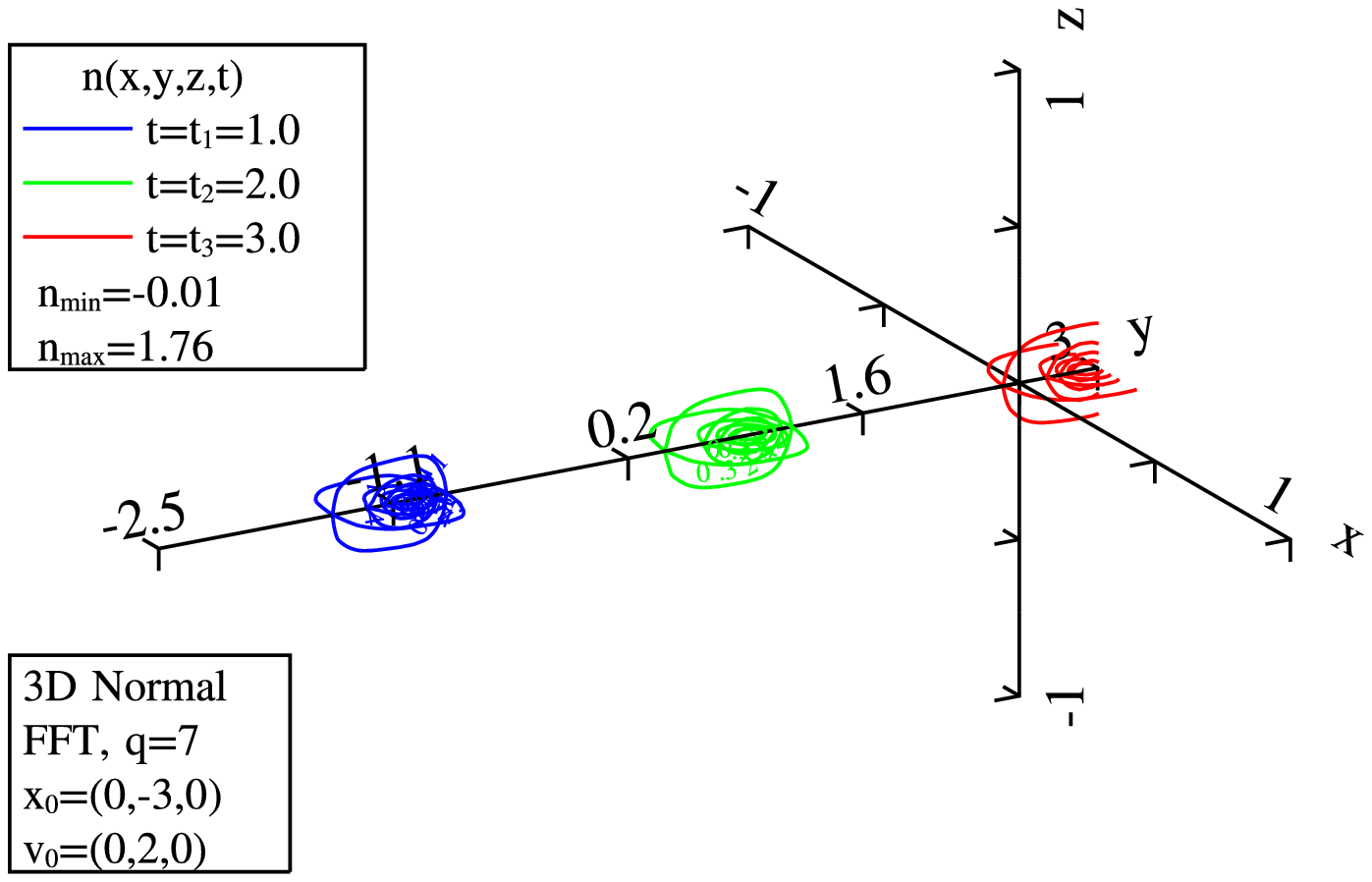}
\caption{\label{3d}The density $n(\vt{x},\st{t})$ for the KV, WB, normal, and Cauchy distributions in 3D space.}
\end{figure*}
 %------------------------------------------------
\section{Numerical methods and results}\label{num}
 %------------------------------------------------
In this section, we briefly discuss the numerical methods we employed and present our results. 
\subsection{A few remarks on integral transforms inversion}
To evaluate the expression (\ref{genfin}) for a particular distribution, we first need to compute $\tlap\tfur_{\vt{k}\st{t}}\left(\st{t} \st{f}_0\left(\vt{v}\right)\right)$ via the formulas presented in the previous section; the expressions are either elementary functions or include special functions or a one-dimensional integral, all these things can be computed straightforwardly. Next step is an evaluation of the inverse Fourier and Laplace transforms. It is well-known that the inverse Fourier transform can be approximated by the discreet Fourier transform and then computed using the FFT algorithm, for details, see Appendix \ref{apena}. In this algorithm, the domain of interest of the resulting function is divided into $N=2^q$ segments.
%, the $q$ value used for computations is shown on the legend of the every plot.
%, we found that for domains we consider $q=10$ gives sufficient accuracy. 
The inverse Laplace transform can be expressed via the inverse Fourier transform:
\begin{align}
\mathcal{L}^{-1}\tilde{f}\left(s\right)&=\frac{\mathrm{e}^{\sigma t}}{2\pi}\int\limits_{-\infty}^{\infty}\tilde{f}\left(\sigma+ik\right)
\mathrm{e}^{ikt}dk\\&=\mathrm{e}^{\sigma t}\mathcal{F}_k^{-1}\tilde{f}\left(
\sigma+ik\right),
\end{align}
which can be evaluated in a way that we described above, 
 $\sigma$ is a real constant greater than the real parts of all singularities of $\tilde{f}$. In the upcoming section, we graphically present our 
results using the dimensionless units. We note that the dimensionless values for the different distributions are not always comparable to each other, since the values for $\vrms$ can be different;
the corresponding conversion factors should be applied.
\subsection{Results and discussion}\label{res}
In this section, we discuss the results obtained numerically and shown in Fig. \ref{1d}, \ref{1dkv}, \ref{2d}, and \ref{3d}; the velocity of the external charge, $\st{v}_0$, is measured in units of $\vrms$ corresponding to the electron's equilibrium distribution and the initial position of the charge,  $\st{x}_0$, is measured in units of $\rd$. The possible space-time ranges differ for the different distributions and are limited by the required precision and the number of points $N=2^q$ in the FFT algorithm. For each plot, we increased $q$ until the values stabilized; the values used are shown in the legends in each plot. The most well-behaving case corresponds to the Cauchy distribution, the KV and normal distributions require greater values of $q$.
For the 1D Cauchy distribution, the numerical results were already shown in Fig. \ref{comp}. \\

In Fig. \ref{1d}, we show the densities computed numerically for all 1D distributions for $\st{v}_0=1$. For the 1D KV distribution, we see that beam's response is a delta function-like peak, for the WB, the density is very spiked and asymmetric. For all distributions, excepting the KV, an accumulation of the charge of the same sign as the external charge occurs. For the normal equilibrium distribution, the perturbation is skewed and spiked resembling the shape of the $\alpha$-stable distribution. For the Cauchy case, the perturbation is also skewed and spiked for $\st{v}_0\approx10^{-1}$, as illustrated in Fig. \ref{comp}. While all other distributions exhibit a symmetric peak around the external charge for $\st{v}_0=0$, the KV distribution has two peaks that spread out with time, as shown in Fig. \ref{1dkv}; with the increase of the velocity, the relative sizes of the peaks change, and, for $\st{v}_0=1$, the left peak almost disappears and the right one looks almost like delta-function, as evidenced in Fig. \ref{1d}.

In Fig. \ref{2d}, the lines of equal densities, for a certain set of times, for all 2D distributions considered, are shown for $\st{v}_0=4$. For the 2D KV distribution, we see the spreading out delta function-like "fronts", similar to the 1D case for $\st{v}_0=0$. For the 2D WB distribution, the lines are triangular with a peak following the charge. For the 2D Cauchy distribution, outer lines are almost circular; for the normal distribution, they have a bit more complicated shape. For the smaller velocities, the profiles are less directed toward the charge.

In Fig. \ref{3d}, the lines of equal densities in certain planes, i.e., in three planes, each of which is parallel to the two out of the three coordinate axes, are illustrated. The shape of the lines for every distribution has the same features as the ones for 1D and 2D cases. 

We emphasize that our 1D and 2D cases are not simple reductions of the 3D case, but correspond to the 1D and 2D theories of electrodynamics with the different from the 3D case Green's functions; the expressions for $\tlap\tfur_{\vt{k}\st{t}}\left(\st{t}\st{f}_0\left(\vt{v}\right)\right)$ also have different forms for the same distribution, but for the different dimensionalities. Although, in the method we consider in the present article, the Green's functions are not used and the solutions of the Poisson equation in the Fourier space were employed, which have the same form (\ref{furim}) in all cases, the general method for a finite beam \cite{Elizarov2012} uses Green's function explicitly and its singularities is the main difficulty there. For the 1D case, the Green's function is not singular and the general method can be implemented without accounting for the singularities and the results can be compared with the exact solution (\ref{huge3}) for the infinite beam; after a certain change of variables, the general method is also able to deal with the infinite beam case. This will be a reliable test of the method itself; and the 2D and 3D results of the present paper will help to develop and test good ways to handle the singularities. The general method is currently under development and we will elaborate on this in our future publications. The similarity of the 1D, 2D and 3D cases is also a sign of the  applicability of the exact  1D results for estimating the shielding in the real device.
\subsection{The code}
The method we discussed herein was implemented as an object-oriented program in C++. The solution is stored as a multidimensional array over some grid in space-time; for further usage, it can be evaluated for any point using interpolation. The program is easily expendable for other external charge densities and equilibrium distributions and, in particular, can deal with the empirical ones. The visualization is also very flexible: it is possible to adjust time values, the number of equal density lines, set the particular values of interest, and look at different projections and cross-sections of the 2D and 3D densities.
\subsection{Application to the Proof-of-principle experiment}
As it was mentioned in the introduction, the proof-of-principle (PoP) experiment is planned at Brookhaven National Laboratory and the corresponding facility is currently under construction \cite{Pinyaev2012}. In this subsection, we describe how the results can be applied to the modulator of the real device. To recover the dimensional quantities, we need the Debye radius and the plasma frequency; for the PoP, we have $\rd=4.65\cdot10^{-5}\,\mathrm{m}$ and $\pf=6.436\cdot 10^9\,\mathrm{s}^{-1}$. We obtain for the dimensional density perturbation:
\begin{align}
n_1\left(\vec{x},t\right)=\frac{1}{\rd^d}\st{n}_1\left(\frac{1}{\rd}\vec{x},\pf t\right),
\end{align}
where $d$ is a spatial dimension of the problem, for the real 3D case, $d=3$. For the PoP experiment, the modulator is constructed such that the interaction time is about one half of the plasma period, it depends on the hadron's velocity, as the modulator length is constant. The velocity is measured in units of $\vrms$, in the PoP experiment, we have $\vrms=3.0\cdot 10^5\,\mathrm{\frac{m}{s}}$. Our computations, shown in Fig. \ref{1d}, demonstrate that extending the modulator up to a few plasma oscillations can significantly increase the density perturbation, i.e., its maximum will be up to four times greater. Further increasing of the modulation time doesn't increase the perturbation, as the modulator saturates, as shown in Fig. \ref{1d}. The amplification of the perturbation in the FEL section is limited by the FEL saturation. For the model-independent description of the FEL saturation and its application to the theory of CeC, see \cite{litv_fel} and \cite{newprl}, respectively. These considerations provide limitations on a possible amplified perturbation that we can get, which, in their turn, determine the performance of the CeC device.
\section{Conclusion and future plans}In the present article, we considered a possible way to model the modulator section of the coherent electron cooling, i.e., we developed a method for evaluating the dynamical shielding of an external charge in an infinite electron plasma; for the certain case, we found analytical solution.
The software package we developed gives reliable results for a variety of equilibrium distributions and initial conditions. We plan to use it in the analysis of the next section of the CeC, the FEL section, wherein the electron density perturbation from the modulator evolves in a free electron laser \cite{Webb, Elizarov2012_2}. The results obtained can also be used as a testing ground for the more general method for a finite beam and for the PIC simulations. \\

Recently, we proposed a full theoretical model of the CeC \cite{Elizarov2012_2}. All sections, i.e., the modulator, the FEL amplifier, and the kicker, were described using the inverse integral transforms. The kicker can be described in a very similar way to the modulator.
%, the difference is that as initial condition we have not only the hadron's charge, but also the amplified perturbation from the FEL section, 
For the details on the FEL section, we refer to \cite{Webb, Elizarov2012_2}. The methods for the inverse integral transforms inversion that we developed and tested in the present article open an opportunity to implement this model and get a reliable and fast complete numerical model of the CeC.

%\newpage %Just because of unusual number of tables stacked at end
%\bibliographystyle{abbrvnat}
%\bibliography{prstab_article}% Produces the bibliography via BibTeX.
\section{Acknowledgments}
Various communications with G. Wang, S. Webb, I. Pogorelov, and A. Fedotov are gratefully acknowledged. We thank A. Woodhead for proofreading.

%------------------------------------------------
\appendix
%------------------------------------------------
\section{Numerical evaluation of the inverse Laplace and Fourier transforms}\label{apena}
\subsection{The Fourier transform}
%------------------------------------------------
We use the following definition of the Fourier transform:
\begin{align}
\mathcal{F}f\left(x\right)\equiv\tilde{f}\left(k\right)=\int\limits_{-\infty}^\infty f\left(x\right)\mathrm{e}^{-ikx}dx,
\end{align}
and the inverse Fourier transform:
\begin{align}
\mathcal{F}^{-1}\tilde{f}\left(k\right)\equiv f\left(x\right)=\frac{1}{2\pi}\int \limits_{-\infty}^\infty\tilde{f}\left(k\right)\mathrm{e}^{ikx}dk.
\end{align}
When we write the Fourier transform or its inverse for the dimensionless variables, we use $\tfur$ and $\tfur^{-1}$, respectively.
The discrete Fourier transform (DFT) assigns to the set of points $\{x_n\}_{0\leq n\leq N-1}$ the set of points $\{X_k\}_{0\leq k\leq N-1}$:
\begin{align}
X_k=\sum\limits_{n=0}^{N-1}x_n\mathrm{e}^{-2\pi i \frac{kn}{N}},
\end{align}
we use the following notation:
\begin{align}
X_k=\mathrm{DFT}_k\left[\{x_n\}_{0\leq n\leq N-1}\right].
\end{align}
The DFT can be computed numerically using the effective fast Fourier transform (FFT) algorithm, there are parallel algorithms allowing to increase the speed of computations.
%------------------------------------------------
\subsection{The Laplace transform}
%------------------------------------------------
The Laplace transform is defined as follows:
\begin{align}
\lap f(t)\equiv \tilde{f}(s)=\int\limits_0^\infty f(t)\e{-ts}dt,
\end{align}
and the inverse:
\begin{align}\label{invlap}
\lap^{-1} \tilde{f}(s)\equiv f(t)=\frac{1}{2\pi i}\int\limits_{\sigma-i\infty}^{\sigma+i\infty} \tilde{f}(s)\e{ts}ds,
\end{align}
where $\sigma$ is a real constant greater than the real parts of all singularities of $\tilde{f}(s)$, we use $\tlap$ and $\tlap^{-1}$ for the dimensionless variables.
\subsection{The inverse Fourier transform via the DFT}
%------------------------------------------------
The inverse Fourier transform can be approximated by the discrete one. Approximating the integral by
\begin{align}
f\left(x\right)\approx\frac{1}{2\pi}\int\limits_{-a}^a \tilde{f}\left(k\right)\mathrm{e}^{ikx}dk,
\end{align}
and, introducing the following notations
\begin{align}
k_{n}=-a+\frac{2a}{N}n,\:x_j=\frac{N\pi}{2a}-\frac{\pi}{a}j, \\C_{j}=\frac{a}{N\pi}\mathrm{e}^{-i\pi\frac{N}{2}}\left(-1\right)^j,
\end{align}
we obtain: 
\begin{align}
f\left(x_j\right)\approx C_{j}\sum\limits_{n=0}^{N-1}\tilde{f}\left(k_{n}\right)\left(-1\right)^n\mathrm{e}^{-i\frac{2\pi nj}{N}} \\ =C_{j}\mathrm{DFT}_j\left[\{\tilde{f}\left(k_{n}\right)\left(-1\right)^n\}_{0\leq n\leq N-1}\right].\label{invfur}
\end{align}
Similar expressions can be written for the multidimensional Fourier transform.
%------------------------------------------------
\subsection{The inverse Laplace transform via the DFT}
%------------------------------------------------
The inverse Laplace transform can be expressed via the Fourier transform:
\begin{align}
\mathcal{L}^{-1}\tilde{f}\left(s\right)=\frac{\mathrm{e}^{\sigma t}}{2\pi}\int\limits_{-\infty}^{\infty}\tilde{f}\left(\sigma+ik\right)
\mathrm{e}^{ikt}dk \\ =\mathrm{e}^{\sigma t}\mathcal{F}_k^{-1}\tilde{f}\left(
\sigma+ik\right),
\end{align}
where $\sigma$ is the same as in (\ref{invlap}) and subscript $k$ stands for the fact that the transform is over $k$. Evaluating the inverse Fourier transform via (\ref{invfur}), we obtain:
\begin{align}
f\left(t_j\right)=\mathrm{e}^{\sigma t_j}
C_{j}\mathrm{DFT}_j\left[\{\tilde{f}\left(
\sigma+ik_{n}\right)\left(-1\right)^n\}_{0\leq n\leq N-1}\right].
\end{align}
%------------------------------------------------
\subsection{The inverse Laplace-Fourier transform via the DFT}
%------------------------------------------------
For the 1D problems, the inverse Laplace-Fourier transform can be computed via the DFT as follows:
\begin{widetext}
\begin{align}
f\left(x_i,t_j\right)=&C_{{i}}C_{{j}}\mathrm{e}^{\sigma t_j}\mathrm{DFT}_{i}\Big[\{
\mathrm{DFT}_{j}\left[\{\tilde{f}\left(
k_{n^{\left(1\right)}},\sigma+i
k_{n^{\left(2\right)}}\right)\left(-1\right)^{n^{\left(2\right)}}\}_{0\leq n^{\left(2\right)}\leq N^{\left(2\right)}-1}\right]\left(-1\right)^{n^{\left(1\right)}}\}_{0\leq n^{\left(1\right)}\leq N^{\left(1\right)}-1}\Big],
\end{align}
\end{widetext}
where superscript $^{\left(2\right)}$ corresponds to the Laplace transform, and $^{\left(1\right)}$ to the Fourier one. For the 2D and 3D cases, the similar expressions can be easily written.
%------------------------------------------------
\section{Special Functions}
%------------------------------------------------
\subsection{The exponential integral functions}
\label{eiig}
%------------------------------------------------
The exponential integral functions are defined by \cite{as}:
\begin{align}
\re{Ei}\left(z\right)=-\int\limits_{-z}^\infty\frac{\e{-t}}{t}dt,\:\:\:\:\: \re{E}_1\left(z\right)=\int\limits_z^\infty\frac{\e{-t}}{t}dt,
\end{align}
these functions
can be computed via the convergent series for a small argument: 
\begin{align}
\re{E}_1\left(z\right)=-\gamma-\ln z+\sum\limits_{k=1}^{\infty}\frac{(-1)^{k+1}z^k}{kk!},\\\re{Ei}\left(z\right)=\gamma+\ln z+\sum\limits_{k=1}^{\infty}\frac{z^k}{kk!},
\end{align}
where $\gamma=0.57721\dots$ is the Euler–-Mascheroni constant,
and via the asymptotic series for a large one:
\begin{align}
\re{E}_1\left(z\right)=\frac{\e{-z}}{z}\sum\limits_{k=0}^{N-1}\frac{k!}{(-z)^n}+O\left(N!z^{-N}\right),
\end{align}
\begin{align}
\re{Ei}\left(z\right)=\frac{\e{z}}{z}\sum\limits_{k=0}^{N-1}\frac{k!}{z^n}+O\left(N!z^{-N}\right),
\end{align}
where it is assumed that  $|\mathrm{arg}z|<\pi$.
%------------------------------------------------
\subsection{The error function}
\label{erf}
The error function and the complementary error function are defined by \cite{as}:
\begin{align}
\mathrm{Erf}(z)=\frac{2}{\sqrt{\pi}}\int\limits_0^x\e{-t^2}dt,\:\:\:\: \mathrm{Erfc}(z)=1-\mathrm{Erf}(z),
\end{align}
respectively,
these functions
can be computed via the convergent series for a small argument and via the asymptotic series for a large one:
\begin{align}
\mathrm{Erf}(z)= \frac{2}{\sqrt{\pi}}\sum\limits_{n=0}^\infty\frac{z}{2n+1}\prod\limits_{k=1}^{n}\frac{-z^2}{k}  , \\
\mathrm{Erf}(z)=1-\frac{\e{-z^2}}{z\sqrt{\pi}}\sum\limits_{n=0}^{N-1}(-1)^n\frac{(2n-1)!!}{\left(2z^2\right)^n}+\notag \\ +O\left(z^{-2N+1}\e{-z^2}\right).
\end{align}
To compute the density perturbation for the normal equilibrium distribution, we need to compute the following expression:
\begin{align}\label{expr}
\tlap\tfur_{\vt{k}\st{t}}\left(\st{t} \st{f}_0\left(\vt{v}\right)\right)=\frac{2}{\st{k}^2}
\left[1-\sqrt{\pi}\e{\st{z}^2}\st{z}\re{Erfc}(\st{z})
\right],
\end{align}
 where $\st{z}=\frac{\st{s}}{|\st{k}|}$.
For certain values of $z$, $\mathrm{Erfc}(z) $ diverges, while $\e{z^2}z\mathrm{Erfc}(z)$ is finite, thus we use the series expansions for the whole expression:
\begin{align}
\e{z^2}z\mathrm{Erfc}(z)= \e{z^2}z\left(1-\frac{2}{\sqrt{\pi}}\sum\limits_{n=0}^\infty\frac{z}{2n+1}\prod\limits_{k=1}^{n}\frac{-z^2}{k}  \right), 
\end{align}
\begin{align}
\e{z^2}z\mathrm{Erfc}(z)=\frac{1}{\sqrt{\pi}}\sum\limits_{n=0}^{N-1}(-1)^n\frac{(2n-1)!!}{\left(2z^2\right)^n}+O\left(z^{-2N+2}\right).
\end{align}
To perform the FFT, one needs to compute the expression (\ref{expr}) exactly at $\st{k}=0$, for this case, we write the following series:
\begin{align}
\tlap\tfur_{\vt{k}\st{t}}\left(\st{t} \st{f}_0\left(\vt{v}\right)\right)=\frac{1}{\st{s}^2}-\frac{1}{\st{s}^2}\sum\limits_{n=2}^{N-1}(-1)^n\frac{(2n-1)!!}
{\left(2\st{z}^2\right)^{n-1}}+\notag \\ +O\left(\st{z}^{-2N+4}\right),
\end{align}
thus
\begin{align}\left.\tlap\tfur_{\vt{k}\st{t}}\left(\st{t} \st{f}_0\left(\vt{v}\right)\right)\right|_{\st{k}=0}=\frac{1}{\st{s}^2}.
\end{align}

\end{document}